\documentclass[aps,pra,reprint,showpacs,superscriptaddress]{revtex4-1}

\usepackage[utf8]{inputenc}
\usepackage{amsmath}
\usepackage{bm}
\usepackage{graphicx}
\usepackage{siunitx}
\usepackage{hyperref}
\newcommand{\braket}[1]{\ensuremath{\langle #1 \rangle}}
\newcommand{\tref}[1]{table~\ref{#1}}
\newcommand{\fref}[1]{figure~\ref{#1}}
\newcommand{\Fref}[1]{Figure~\ref{#1}}

\begin{document}
\title{Probing superfluidity in a quasi two-dimensional Bose gas through its local dynamics}
\author{Camilla De Rossi}
\author{Romain Dubessy}
\author{Karina Merloti}
\author{Mathieu de Go\"er de Herve}
\affiliation{Universit\'e Paris 13, Sorbonne Paris Cit\'e, Laboratoire de physique des lasers, F-93430, Villetaneuse, France}
\affiliation{CNRS, UMR 7538, F-93430, Villetaneuse, France}
\author{Thomas Badr}
\author{Aur\'elien Perrin}
\affiliation{CNRS, UMR 7538, F-93430, Villetaneuse, France}
\affiliation{Universit\'e Paris 13, Sorbonne Paris Cit\'e, Laboratoire de physique des lasers, F-93430, Villetaneuse, France}
\author{Laurent Longchambon}
\affiliation{Universit\'e Paris 13, Sorbonne Paris Cit\'e, Laboratoire de physique des lasers, F-93430, Villetaneuse, France}
\affiliation{CNRS, UMR 7538, F-93430, Villetaneuse, France}
\author{H\'el\`ene Perrin}
\affiliation{CNRS, UMR 7538, F-93430, Villetaneuse, France}
\affiliation{Universit\'e Paris 13, Sorbonne Paris Cit\'e, Laboratoire de physique des lasers, F-93430, Villetaneuse, France}

\begin{abstract}
We report direct evidence of superfluidity in a quasi two-dimensional Bose gas by observing its dynamical response to a collective excitation. Relying on a novel local correlation analysis, we are able to probe inhomogeneous clouds and reveal their local dynamics. We identify in this way the superfluid and thermal phases inside the gas and locate the boundary at which the Berezinskii--Kosterlitz--Thouless crossover occurs. This new analysis also allows to evidence the coupling of the two fluids which induces at finite temperatures damping rates larger than the usual Landau damping.
\end{abstract}
\maketitle

Superfluidity is one of the most fascinating expressions of quantum mechanics at the macroscopic level. It manifests itself through specific dynamical properties, giving rise to collective effects as diverse as quantized vortices or second sound~\cite{Leggett1999}. First observed in liquid helium~\cite{Wilks1987}, superfluidity has also been shown to occur for polaritons in semi-conductor micro-cavities or weakly interacting quantum gases~\cite{Bennemann2013}. Thanks to mature imaging techniques and a high degree of control over their parameters, the latter have proven to be particularly well suited for evidencing key features of superfluids: their linear excitation spectrum at low momentum~\cite{Ozeri2005} as well as the existence of a finite critical velocity have been observed~\cite{Landau1941,Raman1999,Desbuquois2012}. Moreover when set into rotation quantum gases display vortices~\cite{Madison2001,Abo-Shaeer2001} and present specific collective modes~\cite{Stringari1996,
 Guery-Odelin1999}. Yet, extracting relevant information from the dynamics of these systems still represents an experimental challenge.

Properties of quantum gases are fully determined by the knowledge of their density and temperature. Hence, in order to explore different physical regimes, these quantities may have to be varied over a large range of values. While for homogeneous systems each repetition of an experiment will address a single point of this phase space, inhomogeneous quantum gases readily experience different regimes locally. This is the framework of the local density approximation (LDA), which has been extensively exploited while exploring physics of quantum gases \textit{at equilibrium}. It has allowed for instance the determination of the equation of state of a Fermi gas at unitarity~\cite{Nascimbene2010} and of a two-dimensional (2D) Bose gas both in the weakly and strongly interacting regimes~\cite{Rath2010,Yefsah2011a,Ha2013}. In the same spirit, density fluctuations of a quasi-one-dimensional Bose gas have been described from the quasi-condensate to the nearly ideal gas regime~\cite{Esteve2006}.

In this work, we extend this approach to the in situ study of the \textit{dynamical} properties of an inhomogeneous quantum gas. We apply a novel \textit{local correlation analysis} to probe the superfluidity of a quasi-2D Bose gas, observing its response to the so-called scissors excitation~\cite{Guery-Odelin1999,Marago2000}. It allows us to identify and locate the superfluid and normal phases within the system. In addition, at finite temperature we evidence a collisional coupling between them. In a sense, our scheme relying on the weak global excitation of a gas close to equilibrium probed locally is symmetric to the recent measurement of the critical velocity in an inhomogeneous quasi-2D Bose gas, where the effect of a strong local perturbation is probed through a global observable, the total heat deposited over the whole cloud~\cite{Desbuquois2012}.

The reduction to low dimensions strongly affects the physics of the collective properties of a quantum system~\cite{Pricoupenko2003,Bloch2008a}. In particular, in two dimensions, thermal phase fluctuations destroy long range order~\cite{Mermin1966} and prevent the occurrence of Bose-Einstein condensation (BEC) in a homogeneous Bose gas at any finite temperature~\cite{Posazhennikova2006}. On the other hand, these systems undergo a Berezinskii--Kosterlitz--Thouless (BKT) phase transition to a superfluid state~\cite{Berezinskii1971,Kosterlitz1973,Bishop1978}. This still holds for quantum gases confined in harmonic traps~\cite{Holzmann2008a} as evidenced experimentally for quasi-2D Bose gases through the algebraic decay of their phase coherence~\cite{Hadzibabic2006}. Here, we directly locate the boundary where the superfluid fraction vanishes across the BKT crossover  through local correlation analysis.

\begin{figure}[t]
\centering
\includegraphics[scale=1]{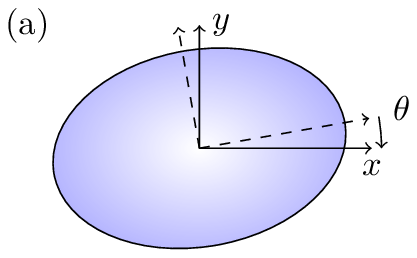}\qquad
\includegraphics[scale=1]{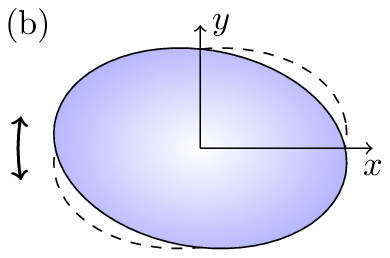}\\
\includegraphics[width=\linewidth]{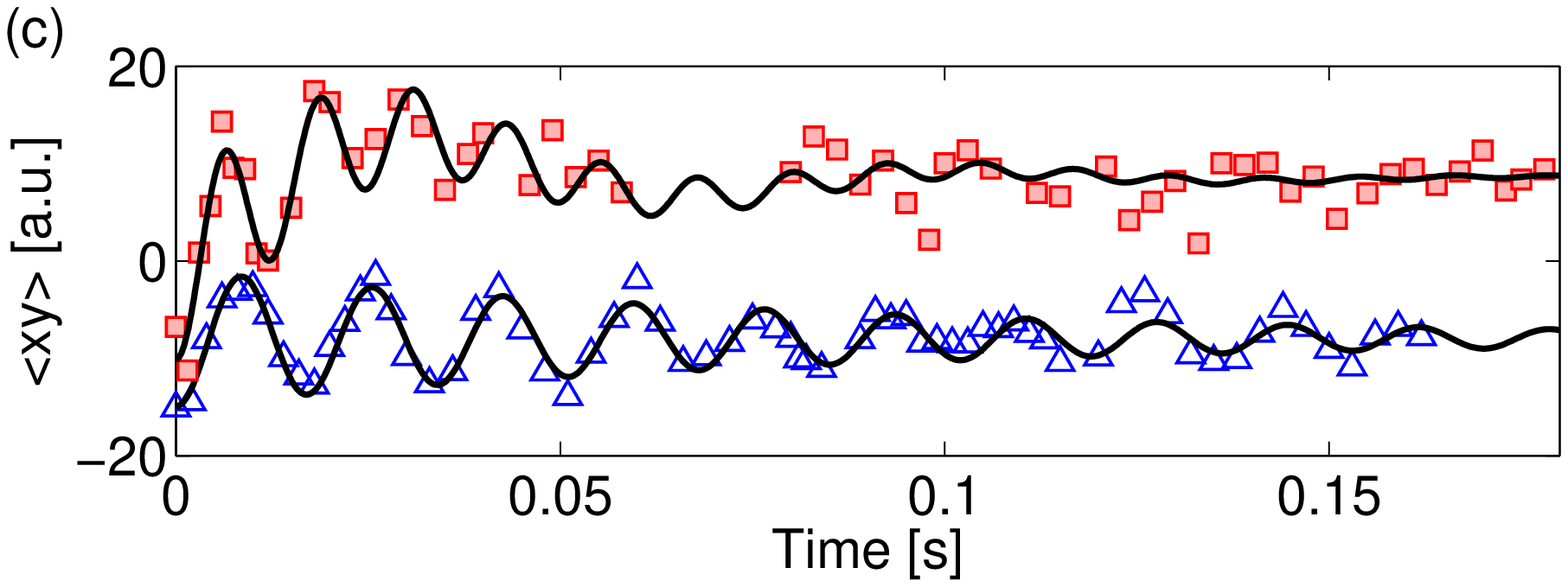}
\caption{\label{fig:experiment}
Principle of the experiment studied in this work.
(a) A quasi-2D sample (blue shaded ellipse) is prepared in an anisotropic harmonic trap (dashed axes), and at $t=0$ the trap axes are rotated by $\theta\simeq\SI{10}{\degree}$.
(b) For $t>0$, the cloud shape oscillates in the rotated trap (solid axes): the scissors mode is excited.
(c) Examples of the evolution with holding time of the average $\braket{xy}$ over the in situ atomic density distribution, for an almost purely degenerate sample ($\alpha=0.78(18)$, open blue triangles) and a non-degenerate sample ($\alpha=0.09(2)$, filled red squares), together with the corresponding fits (black solid lines). Each point is an average of two measurements. A constant offset has been subtracted from the first dataset for the sake of clarity.
}
\end{figure}

The starting point for our experiments is a partially degenerate Bose gas of $^{87}$Rb atoms produced in a radio-frequency (rf) dressed quadrupole trap~\cite{Merloti2013}, loaded from an hybrid trap~\cite{Dubessy2012a}. For our experimental parameters the measured vertical trapping frequency $\omega_z=2\pi\times\SI{1.83\pm0.01}{\kilo\hertz}$ is much larger than the in-plane trapping frequencies $\omega_x=2\pi\times\SI{33.8\pm0.2}{\hertz}$ and $\omega_y=2\pi\times\SI{48.0\pm0.2}{\hertz}$, resulting in a pancake shaped atomic cloud. The trap in-plane geometry, defined by the direction of the eigenaxes $(x,y)$ and the anisotropy $\epsilon=(\omega_y^2-\omega_x^2)/(\omega_y^2+\omega_x^2)=0.34(5)$, is controlled by the rf polarization.

We prepare different samples by varying the total atom number and temperature. As both the temperature $k_BT$ and the chemical potential $\mu$ are comparable to the vertical oscillator energy level spacing $\hbar\omega_z$, the system is in the quasi-2D regime, and the effective dimensionless 2D interaction constant is $\tilde{g}=\sqrt{8\pi}a/\ell_z=0.1056(3)$~\cite{Petrov2000}, where $a$ is the scattering length, $\ell_z=\sqrt{\hbar/(m\omega_z)}$ the vertical harmonic oscillator length and $m$ the atomic mass. For most of our datasets the fraction of atoms out of the vertical oscillator groundstate is smaller than $\SI{20}{\percent}$, despite the relatively high temperatures and chemical potentials\footnote{See supplementary material for details on the imaging techniques, data processing and models.\label{Note1}} (with respect to $\hbar\omega_z=h\times\SI{1830}{Hz}=k_B\times\SI{88}{nK}$). Therefore the system is well described by coupled 2D manifolds and exhibits 2D physics. Specifically, the BKT transition is expected to occur in our trap at a phase space density $\mathcal{D}_c=8.188(3)$ for a reduced chemical potential $\alpha = \mu/(k_B T)$ equal to $\alpha_c=\mu/(k_BT)=0.162(1)$~\cite{Prokofev2001}. These critical values increase slightly when the effect of the population in the transverse excited states at $k_BT\gtrsim \hbar\omega_z$ is taken into account \cite{Holzmann2010}. The samples are prepared with a reduced chemical potential $\alpha$ in the range $[0.09-0.88]$, on both sides of the expected threshold \cite{Note1}.

We excite the scissors oscillations by suddenly changing the orientation of the horizontal confinement axes by $\theta \simeq \SI{10}{\degree}$, while keeping the trap frequencies constant~\cite{Dubessy2014}, see \fref{fig:experiment}(a). The superfluid and thermal fractions of the cloud are subsequently expected to oscillate at different frequencies: $\omega_\pm=|\omega_x\pm\omega_y|$ for a normal gas close to the collisionless regime and $\omega_{\rm hd}=\sqrt{\omega_x^2+\omega_y^2}$ for the superfluid~\cite{Guery-Odelin1999}. We typically record one hundred in situ pictures of the atoms after different holding times, spanning about $\SI{200}{\milli\second}$. In order to study the scissors mode at finite temperature, it is essential to be able to discriminate between the superfluid and thermal phases~\cite{Marago2001}.  However a direct fit of the 2D in situ density profiles by a bi-modal function is quite imprecise because the typical width over which they extend are comparable.

Inspired by the theoretical works which use the average $\braket{xy}$ as an observable of the scissors mode excitation~\cite{Guery-Odelin1999,Jackson2001,Simula2008}, we use an efficient algorithm to extract $\braket{xy}$ from our data \cite{Note1}. For all of our datasets $\braket{xy}$ presents damped oscillations as illustrated in \fref{fig:experiment}(c). In order to quantitatively analyze this signal, we take advantage of the classical description of the scissors oscillations~\cite{Guery-Odelin1999}. In this framework the complex oscillation frequencies $\omega$ of $\braket{xy}$ are parametrized by a characteristic relaxation time $\tau$:
\begin{equation}
i\frac{\omega}{\tau}(\omega^2-\omega_{\rm hd}^2)+(\omega^2-\omega_+^2)(\omega^2-\omega_-^2)=0.
\label{eqn:hydro}
\end{equation}
In the hydrodynamic limit where $\tau$ vanishes, this model also describes the scissors mode in a superfluid~\cite{Guery-Odelin1999}. Applying this prediction to our data, we are able to fit \cite{Note1} the value of $\tau$ and deduce the frequencies $\omega_{\rm sc}$ and damping rates $\Gamma_{\rm sc}$ of the oscillations relying on the constraints of equation~\eqref{eqn:hydro}. In this way, we perform a double frequency fit with only three parameters (amplitude, offset and $\tau$). The model always predicts two frequencies, corresponding to an upper branch and a lower branch. The former spans frequencies in the range $[\omega_{\rm hd},\omega_+]$ while the latter spans $[0,\omega_-]$. We found this method to be more accurate than a fit to a sum of two damped cosines which requires at least nine free parameters, in particular for the determination of the lowest, highly damped, frequency. We checked that the fit constrained by equation~\eqref{eqn:hydro} introduces no bias and that both models give the same results \cite{Note1}.
\begin{figure}[t]
\centering
\includegraphics[width=\linewidth]{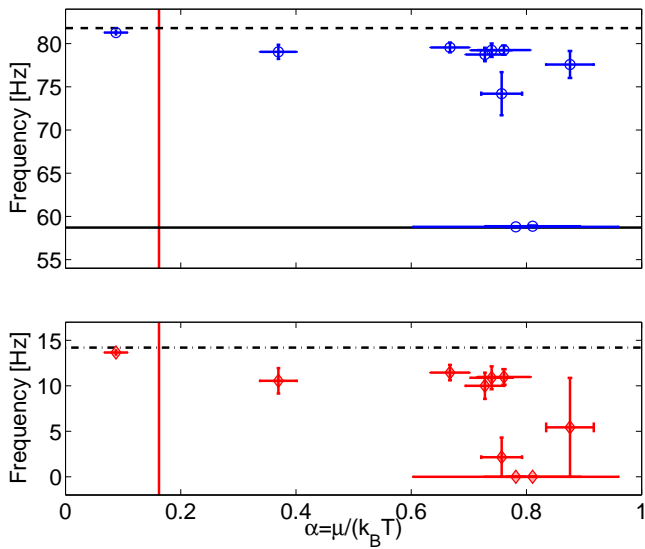}
\caption{\label{fig:frequencies_whole}
Measured oscillation frequencies as a function of the reduced chemical potential $\alpha=\mu/(k_BT)$. The open blue circles correspond to the upper frequency branch, the filled red diamonds to the lower frequency branch, see text for detail. The three horizontal lines indicate the expected frequencies in the low/high temperature limits: $\omega_{\rm hd}/(2\pi)$ (solid line), $\omega_+/(2\pi)$ (dashed) and $\omega_-/(2\pi)$ (dashed--dotted). The vertical red solid line indicates the estimated position of the BKT transition $\alpha_c=0.162(1)$ for a 2D Bose gas with our trap parameters~\cite{Prokofev2001}. The error bars are the standard deviations estimated using the fit residuals.
}
\end{figure}

\Fref{fig:frequencies_whole} reports the frequencies found as a function of the reduced chemical potential $\alpha$. Below the BKT transition ($\alpha<\alpha_c$, leftmost point of the graph), the two frequencies are very close to, but slightly below, the classical prediction $\omega_\pm$. This, together with the large observed damping rates  (see \fref{fig:experiment}(c)), indicates that the gas is not fully in the collisionless regime, which would require $\omega_{\rm sc}\tau\gg 1$~\cite{Guery-Odelin1999}. For samples above the BKT transition ($\alpha>\alpha_c$) we find either a single frequency close to $\omega_{\rm hd}$ or two frequencies, shifted below the collisionless frequencies $\omega_\pm$. Our data are consistent with an increase of the upper frequency from $\omega_{\rm hd}$ to $\omega_+$ when $\alpha$ decreases, in agreement with 2D dynamical classical field simulations~\cite{Simula2008}, and in stark contrast with what was observed in three-dimensional experiments~\cite{Marago2001}. We note however that the simple criterion based on the critical value $\alpha_c$ is not sufficient to describe our datasets: in particular the thermal frequencies $\omega_\pm$ are still observable beyond $\alpha=\alpha_c$. The global $\braket{xy}$ observable therefore fails to evidence the normal to superfluid crossover. Indeed, in these conditions the superfluid and normal phases coexist in the trap and the simple picture of equation~\eqref{eqn:hydro} does not capture all the dynamics. Moreover, as the observable $\braket{xy}$ is computed over the whole sample, both phases are combined in the total signal and it is difficult to isolate their respective signature.
\begin{figure}[t]
\centering
\includegraphics[width=\linewidth]{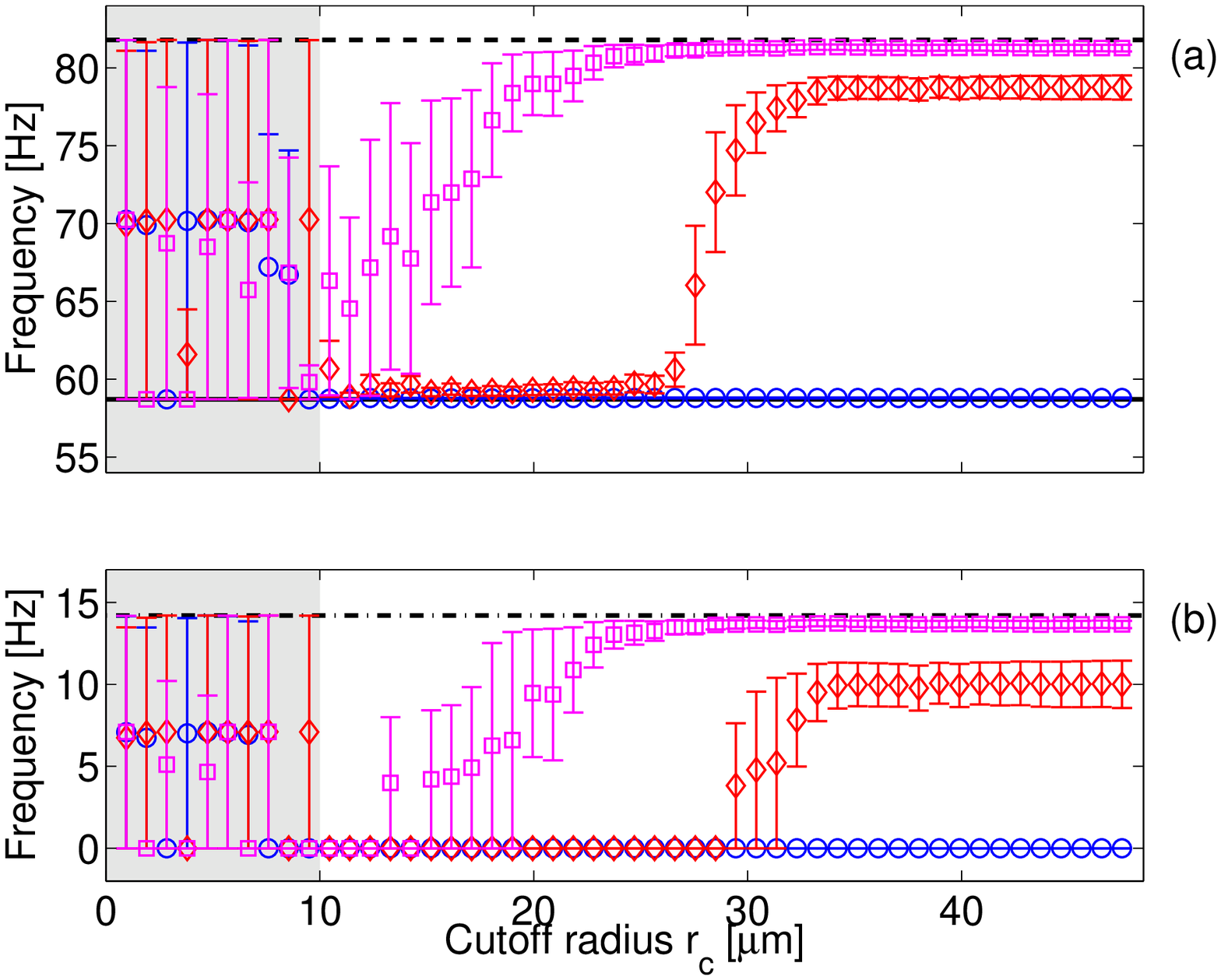}\\
\includegraphics[scale=1]{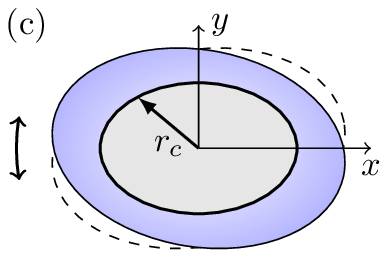}
\caption{\label{fig:frequency_part}
Measured oscillation frequencies of the $\braket{xy}_{r_c}$ observable as a function of the cutoff radius $r_c$ (see text). (a) corresponds to the upper branch, (b) to the lower branch. Measurements for three different experimental conditions are presented: $\alpha=0.78(18)$ (open blue circles), $\alpha=0.73(3)$ (open red diamonds) and $\alpha=0.09(2)$ (open magenta squares). The three horizontal solid lines correspond to the three expected frequencies, as in \fref{fig:frequencies_whole}. The gray shaded areas indicate the central part of the clouds where the signal to noise ratio is too small to extract a frequency for all the datasets. The error bars are the standard deviations estimated using the fit residuals. (c) The computation of $\braket{xy}_{r_c}$ uses only the pixels located at a scaled radius $\bar{r}<r_c$ (gray shaded ellipse).}
\end{figure}

In order to overcome this issue we introduce the concept of \emph{local correlation analysis}. In the spirit of LDA, we define the rescaled radius $\bar{r}=\sqrt{(\omega_x/\omega_y)x^2+(\omega_y/\omega_x)y^2}$, following a 2D isodensity path around the trap center. We compute a partial average $\braket{xy}_{r_c}$ evaluated by including only the pixels at a rescaled distance $\bar{r}<r_c$. In this way we expect to isolate the superfluid phase which is located at smaller $\bar{r}$.

\Fref{fig:frequency_part} presents the frequencies found as a function of the cut-off radius $r_c$, for three of the datasets. Below a cutoff radius of $r_c\simeq\SI{10}{\micro\meter}$ there is no clear oscillation (indicated by the large error bars). This reflects the fact that the scissors excitation is a surface mode and does not affect the core of the cloud. Beyond this cutoff radius, we observe in most of the datasets a central region oscillating at a single frequency, in good agreement with the hydrodynamic prediction $\omega_{\rm hd}$, which was not apparent in the global analysis. For larger radii, we recover the results of the global analysis, showing that the contribution of the central region to the total signal is small, even if the density is higher at the center. The radius at which the crossover between these two regimes occurs depends on $\alpha$. As discussed below, this result is a direct evidence of the coexistence of normal and superfluid scissors excitations at finite temperature in our system.

\begin{figure}[t]
\centering
\includegraphics[width=\linewidth]{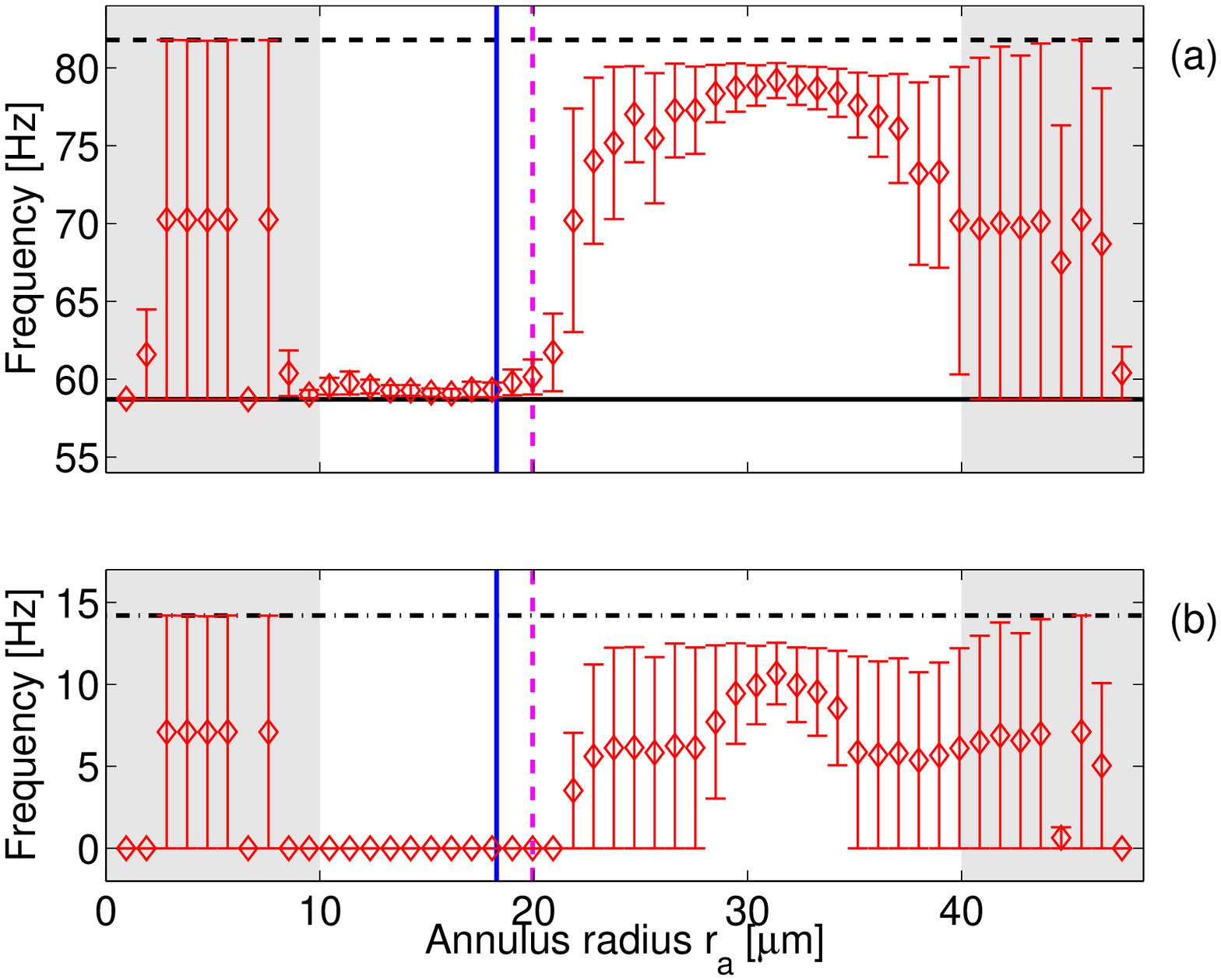}\\
\includegraphics[scale=1]{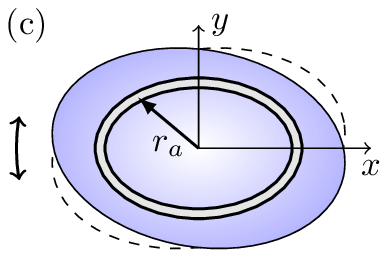}
\caption{\label{fig:frequency_annulus}
Measured frequencies for $\braket{xy}_{r_a}$, computed on a thin annulus of width $\delta r=4$ pixels, centered on a given scaled radius $\bar{r}=r_a$, as sketched in (c). (a) corresponds to the upper branch, (b) to the lower branch. The error bars are the standard deviations estimated using the fit residuals. The grey shaded areas indicate the low signal to noise ratio regions, at the cloud center (left) and in the wings (right). The two vertical lines are estimates of the boundary between the superfluid and normal phases in a trapped quasi-2D Bose gas, based on the local density approximation~\cite{Holzmann2010} (solid blue), and on the local collision time (dashed magenta), see text for details.}
\end{figure}

To get more accurate results, we reduce further the region of interest to a thin annulus centered on $\bar{r}=r_a$, with a width $\delta r=4$ pixels, thus probing an isodensity region of the cloud. \Fref{fig:frequency_annulus} displays the frequencies measured for this partial average $\braket{xy}_{r_a}$ as a function of $r_a$, for $\alpha=0.73(3)$, corresponding to one of the datasets of \fref{fig:frequency_part}.
The larger error bars result from a smaller number of pixels involved in the average. This local analysis is even more sensitive to frequency shifts and allows a more accurate determination of the position of the boundary between the two phases at $\bar{r}\simeq\SI{22}{\micro\meter}$.

We attribute the hydrodynamic part of the gas present at the center to the superfluid phase of the degenerate quasi-2D Bose gas. One may wonder if a hydrodynamic classical gas at high density could explain our observation. From the independent measurement of the equilibrium values of $\mu$ and $T$, we evaluate the scaled distance from the trap center at which the reduced local chemical potential $\alpha(\bar{r})=\mu_{\rm loc}(\bar{r})/(k_BT)$, where $\mu_{\rm loc}(\bar{r})=\mu-m\omega_x\omega_y\bar{r}^2/2$, reaches the critical value for the BKT transition~\cite{Holzmann2010}. This is indicated by the vertical blue solid line in \fref{fig:frequency_annulus}, in good agreement with our observed crossover. The small discrepancy could be explained in two ways. First,  close to the critical radius, the finite extension $\delta r$ of our local probe averages the signal over both the superfluid and normal regions. Second, finite size effects for a trapped cloud, away from the thermodynamic limit, modify the position of the expected boundary anyway. Indeed, a Quantum Monte Carlo simulation of a quasi two-dimensional trapped Bose gas with our parameters\footnote{Markus Holzmann, private communication.} has shown that the normal to superfluid crossover is broadened by about \SI{2}{\micro\meter} in the sample, resulting in a non-zero superfluid fraction beyond the critical radius predicted by the LDA. We are thus confident that we observe the superfluid to classical transition signature in the dynamical response of the gas.

We now come to the interpretation of the local relaxation time $\tau(r_a)$ deduced from the local analysis. We find that its value is in good agreement with the inverse of the local collision rate $\Gamma_c(r_a)\simeq n_{2D}(r_a)\tilde{g}^2\hbar/m$ evaluated in the confinement dominated three-dimensional regime~\cite{Petrov2001}, where $n_{2D}(r_a)$ is the measured average 2D density on the annulus. This supports the use of equation~\eqref{eqn:hydro} to describe the local dynamics. We point out that our analysis could give access to a local relaxation time of the excitation, a key ingredient in finite temperature two-fluid models~\cite{Williams2001,Nikuni2002}.

Conversely we use the measured $\tau(r_a)$ to estimate the local collision rate and hence the local phase space density, knowing the sample temperature. The vertical dashed magenta line in \fref{fig:frequency_annulus} indicates the radius at which this estimated local phase space density becomes higher than the BKT critical phase space density $\mathcal{D}_c=11.2$ computed for this quasi-2D gas~\cite{Holzmann2010}. The analysis of the dynamics thus confirms the location of the boundary between the superfluid and normal phases.

\begin{figure}[t]
\centering
\includegraphics[width=\linewidth]{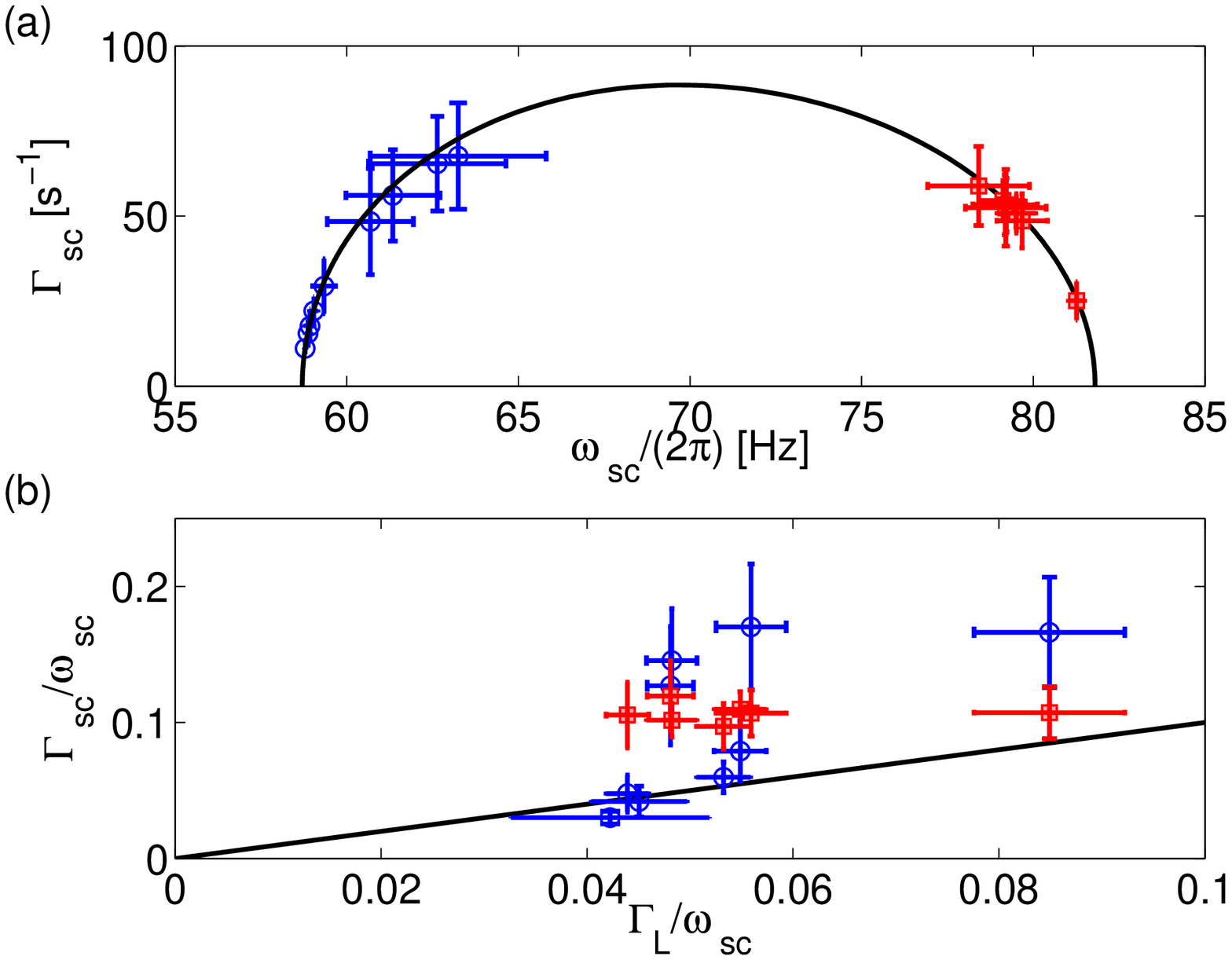}\\
\includegraphics[scale=0.9]{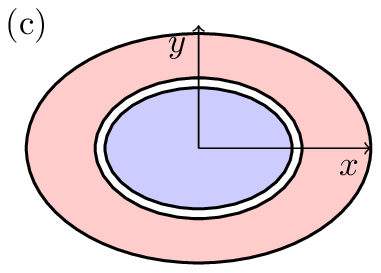}
\caption{\label{fig:damping_Landau}
(a) Measured damping rate and frequencies for the central superfluid phase (open blue circles) and for the normal phase (filled red squares), for all the datasets. The black solid curve is the prediction of equation~\eqref{eqn:hydro}. The error bars are the standard deviations estimated using the fit residuals.
(b) Measured damping rate, in units of the scissors oscillation frequency, for the central superfluid phase (open blue circles) and for the normal phase (filled red squares), as a function of the reduced damping Landau rate of equation~\eqref{eqn:Landau_damping}, estimated for each dataset from the values of $\mu$ and $T$. The black solid line is a guide to the eye (with slope 1). The error bars are the standard deviations estimated using the fit residuals.
(c) Contributions from the superfluid (central blue shaded area) and the normal phase (outer red shaded area) to the computation of $\braket{xy}$ are isolated. The thin excluded area (white annulus) corresponds to the crossover evidenced using the method of~\fref{fig:frequency_annulus}, for each of the datasets.}
\end{figure}

We now turn to a quantitative analysis of the damping rates $\Gamma_{\rm sc}$ of the local correlations. We will only consider the upper frequency branch, sketched on \fref{fig:damping_Landau}(a), as the lower frequency is highly damped in our experiments. In order to get a reliable estimate of the frequency and damping rates of the scissors oscillation for both the central and outer regions, we exclude the crossover area and compute the $\braket{xy}$ averages over a disc for the central part and over a large annulus for the outer part, as illustrated on \fref{fig:damping_Landau}(c). \Fref{fig:damping_Landau}(b) compares the measured reduced damping rate of the scissors oscillations $\Gamma_{\rm sc}/\omega_{\rm sc}$ to the reduced Landau damping $\Gamma_L/\omega_{\rm sc}$~\cite{Pitaevskii1997,Fedichev1998}:
\begin{equation}
\frac{\Gamma_L}{\omega_{\rm sc}}=\frac{3\pi}{8}\frac{k_BTa}{\hbar c},
\label{eqn:Landau_damping}
\end{equation}
where $c=\sqrt{2\mu/(3m)}$ is the sound velocity for a pancake-shaped Bose gas~\cite{DeRosi2015}.
For our coldest samples the measured damping rate for the superfluid phase is close to the damping predicted by equation~\eqref{eqn:Landau_damping}. However for the datasets with a large thermal fraction, the damping in the superfluid phase deviates from the predicted Landau damping and is of the same order of magnitude than the damping in the normal phase, suggesting that this additional damping originates from collisional coupling to the normal gas~\cite{Williams2001}. The reduced damping rate in the normal phase is approximately constant for all our datasets. We note that in our 2D geometry the abnormal damping cannot be explained by a resonant Beliaev coupling to other modes as observed in three-dimensional experiments~\cite{Hodby2001}.

In conclusion we have studied the dynamics of a quasi-2D Bose gas near the BKT transition. We have shown the coexistence of a superfluid fraction and a thermal part by a direct, in situ, local analysis of its dynamical behavior. We use the frequency of the scissors mode as a local probe of superfluidity. Whereas a global analysis of the data fails to reveal the coexistence of the superfluid and normal oscillations, we demonstrate a new local correlation analysis, reminiscent of the local density approximation, which allows to locate the normal to superfluid transition, using a purely dynamical criterion. The position of the boundary is in good agreement with the crossover expected from the equilibrium properties of the gas. In principle it should be possible with this local probe to observe the superfluid density jump at the BKT transition~\cite{Nelson1977,Bishop1978}. This motivates future numerical studies of finite temperature dynamics to establish quantitatively the relation between local scissors frequency and local superfluid fraction.

Local correlation analysis could also be applied to numerical simulations. In particular, the local measurement of relaxation times would be useful to compare experiments to numerical calculations at finite temperature. We expect this new kind of local diagnosis to shed new light on the study of out-of-equilibrium systems.

\begin{acknowledgments}
We thank M. Holzmann for enlightening discussions and for providing us with the QMC simulations of our experiment, S. Stringari for helpful suggestions and acknowledge stimulating discussions with B. Laburthe-Tolra. We thank D. Gu\'ery-Odelin for a critical reading of the manuscript. We acknowledge financial support from ANR project SuperRing (ANR-15-CE30-0012-01). LPL is a member of Institut Francilien de Recherche sur les Atomes Froids (IFRAF).
\end{acknowledgments}

\section*{Supplementary material}
\subsection{Image processing}
The determination of the equilibrium properties of the cloud follows the approach described in reference~\cite{Hadzibabic2008}. Prior to the excitation we take an in situ picture of the atoms at rest in the initial trap using a low intensity ($I\sim 0.7 I_{\rm sat}$) imaging pulse, to recover correctly the low density profile of the thermal wings. We then adjust this density profile by a self consistently determined Hartree-Fock mean field model including up to ten excited states manifolds. From this procedure we deduce the chemical potential $\mu$ and temperature $T$ of the cloud~\cite{Hadzibabic2008}. The total atom number and detection efficiency are calibrated independently using an auxiliary time of flight imaging system observing the atoms at lower density from the side~\cite{Dubessy2012a}. The equilibrium parameters for the different datasets are summarized in \tref{tab:datasets}.
\begin{table}[b]
\caption{\label{tab:datasets}
Equilibrium properties of the cold atom ensembles, prior to the scissors oscillation excitation: chemical potential $\mu$, temperature $T$, reduced chemical potential $\alpha=\mu/(k_BT)$ and ratio of the number of atoms in the groundstate of the vertical oscillator $N_{0z}$ to the total number of atoms $N$. The data plotted on figure 1(c) and 3 of the main paper come from datasets 1 and 9. Dataset 7 is also used for figure 3 and 4.}
\begin{ruledtabular}
\begin{tabular}{ccccc}
Dataset&$\mu/h$ (Hz)&$T$ (nK)& $\alpha=\mu/(k_BT)$ & $N_{0z}/N$\\
1 & $2165\pm380$ & $133\pm7$ & $0.78\pm0.18$ & 0.90\\
2 & $2651\pm190$ & $157\pm5$ & $0.81\pm0.08$ & 0.88\\
3 & $2633\pm68$ & $167\pm4$ & $0.76\pm0.04$ & 0.86\\
4 & $2936\pm84$ & $161\pm3$	 & $0.88\pm0.04$ & 0.88\\
5 & $2057\pm63$ & $148\pm3$ & $0.67\pm0.03$ & 0.86\\
6 & $3081\pm91$ & $200\pm4$ & $0.74\pm0.04$ & 0.82\\
7 & $3169\pm85$ & $209\pm4$ & $0.73\pm0.03$ & 0.81\\
8 & $3598\pm140$ & $227\pm5$ & $0.76\pm0.05$ & 0.80\\
9 & $396\pm86$ & $217\pm5$ & $0.09\pm0.02$ & 0.42\\
10 & $1955\pm130$ & $254\pm5$ & $0.37\pm0.03$ & 0.59
\end{tabular}
\end{ruledtabular}
\end{table}
The number of atoms in the groundstate of the vertical harmonic oscillator $N_{0z}$ is estimated from the values of the chemical potential and temperature using an hybrid Thomas-Fermi plus Hartree-Fock mean field model that allows to compute the populations in the first ten vertical oscillator manifolds. Since a majority of the atoms are in the groundstate the system is in the quasi two-dimensional regime.

After the excitation, we record the atomic density after a variable delay time using a saturating imaging pulse ($I\sim 8 I_{\rm sat}$)~\cite{Reinaudi2007}. We are thus able to resolve both the low density thermal wings of the cloud and the high density degenerate core. Our home-made imaging system has a resolution of $\SI{4}{\micro\meter}$ and an effective pixel size of $\SI{0.95}{\micro\meter}$ in object space.

\begin{figure}[t]
\centering
\includegraphics[width=\linewidth]{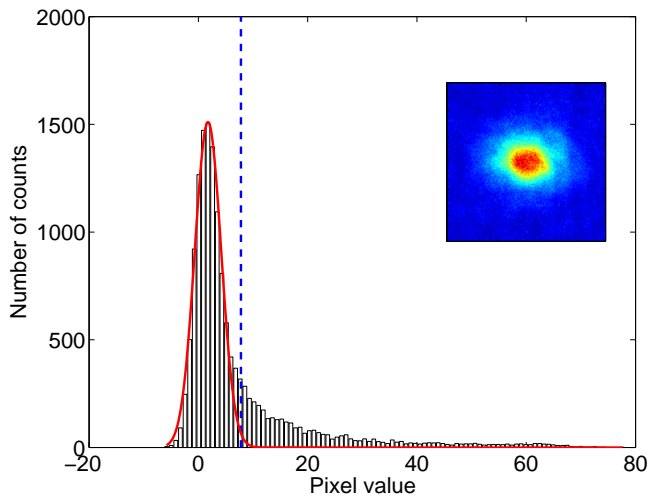}
\caption{\label{app:histogram}
Data histogram (white bars) of a single picture of $121\times 121$ pixels (see inset). The red solid line is a Gaussian fit to the background distribution, and the vertical blue dashed line is the estimated signal threshold.
}
\end{figure}

In order to compute the relevant density weighted average values $\braket{x}$, $\braket{y}$, $\braket{x^2}$, $\braket{y^2}$ and $\braket{xy}$, it is necessary to remove the imaging noise which does not correspond to real atomic signal. We proceed as follows. For a square picture, and a small ellipsoidal cloud, about half the pixels do not represent atoms and therefore contain information only about the noise. This appears clearly if one plots the histogram of the picture, as we do in \fref{app:histogram}.
We identify a large contribution due to the background, with a low value of signal, and a smaller contribution due to the data, at larger signal values. From a Gaussian fit to the background contribution, we estimate a signal threshold ($2.5\sigma$ of the Gaussian) above which we can neglect the noise. We then cancel the contribution of all pixels having a value lower than the threshold. Note that this process also filters a few relevant pixels, but it greatly improves the signal to noise ratio. We checked that the results are insensitive to the exact position of the threshold.

After this noise removal procedure, we compute the values of $\braket{x^\prime}$, $\braket{y^\prime}$, $\braket{x^{\prime 2}}$, $\braket{y^{\prime 2}}$ and $\braket{x^\prime y^\prime}$ for each picture, where $x^\prime$ and $y^\prime$ are the coordinates of the pixels in the camera frame. By adjusting the two-dimensional center of mass motion of the cloud we find the trap frequencies, the trap center, and the orientation of the axes relative to the camera. Then by combining with appropriate weights the second order moments we extract the $\braket{xy}$ values in the trap frame.

\subsection{Scissors mode classical model}
In this section we describe the fitting procedure used in the main paper to extract the frequencies and relaxation times of the scissors oscillations. We start from the following set of coupled equations describing the classical scissors oscillations~\cite{Guery-Odelin1999}:
\begin{subequations}
\begin{eqnarray}
\frac{d\braket{xy}}{dt}&=&\braket{xv_y+yv_x},\\
\frac{d\braket{xv_y+yv_x}}{dt}&=&-(\omega_x^2+\omega_y^2)\braket{xy}+2\braket{v_xv_y},\\
\nonumber\frac{d\braket{v_xv_y}}{dt}&=&-\frac{\omega_x^2+\omega_y^2}{2}\braket{xv_y+yv_x}-\frac{\braket{v_xv_y}}{\tau}\\
&&-\frac{\omega_y^2-\omega_x^2}{2}\braket{yv_x-xv_y},\\
\frac{d\braket{yv_x-xv_y}}{dt}&=&(\omega_y^2-\omega_x^2)\braket{xy}.
\end{eqnarray}
\label{eqn:dyn_xy}
\end{subequations}

\begin{figure}[t]
\includegraphics[width=\linewidth]{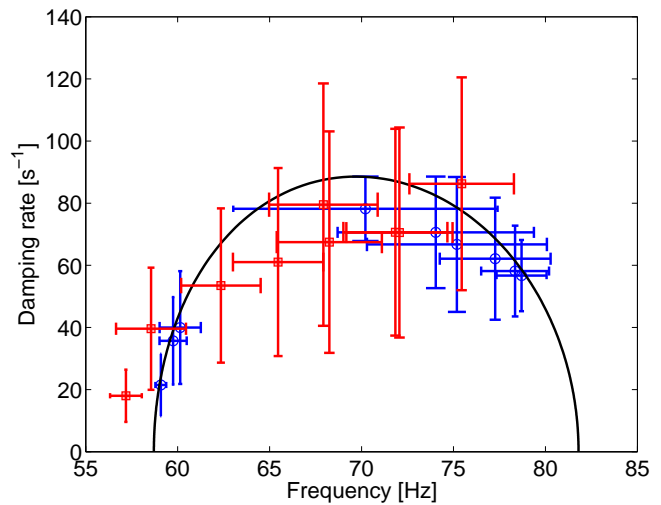}
\caption{\label{fig:freq_damping}
Measured frequencies and damping rates (for the upper branch) using the local $\braket{xy}$ probe and two different fitting methods: open blue circles are obtained using the model derived from equations~\eqref{eqn:dyn_xy}, filled red squares are obtained using a sum of two damped cosines model. The error bars are the standard deviations estimated using the fit residuals. The black solid line is the upper branch predicted using equation~(1). For the sake of clarity only the data points corresponding to radii $r_a=\{12,17,21,23,24,25,28,30,35\}$ are displayed.}
\end{figure}

We denote $f_\tau(t)=\braket{xy}(t)$ the solution of equations~\eqref{eqn:dyn_xy} with initial conditions: $f_\tau(0)=\braket{xy}_0$ (an arbitrary non zero value), $\braket{yv_x\pm xv_y}_{t=0}=0$ and $\braket{v_xv_y}_{t=0}=0$. We use the time-dependent function $g(\tau,A,B): t\mapsto Af_\tau(t)+B$ as a model to fit our experimental data, with $A$, $B$ and $\tau$ as free real parameters. For small values of $\tau$ this model is close to a damped cosine, with frequency $\omega\simeq\omega_{\rm hd}=\sqrt{\omega_x^2+\omega_y^2}$, while for large values of $\tau$ this model is close to a sum of two damped cosines with frequencies close to $\omega_\pm=|\omega_x\pm\omega_y|$. In both cases the damping depends on the exact value of $\tau$. Once we have determined the value of $\tau$ from the fit, we solve equation~(1) of the main paper to find the corresponding frequency and damping rate.

\Fref{fig:freq_damping} displays the damping rates and frequencies found for different radii $r_a$ for the same dataset as in figure~4, using two fitting methods: the one described above and a simple sum of two damped cosines. Both models give results close to the prediction of equation~(1) (considering the error bars). As mentioned before the former model has been derived from the predictions of equation~(1) and therefore the points must fall near the theoretical curve. However the latter model do not make any kind of assumption constraining the relative values of the damping rate and the frequency and yet agrees with the same theoretical prediction. Therefore we conclude that equation~(1) quantitatively describes the scissors oscillations found with our local correlation analysis scheme. Using our model fit enables to reduce the error bars and thus to extract the low frequency branch from the noise.

%

\end{document}